\documentclass[journal]{IEEEtran}
\usepackage{amsmath}
\usepackage{amssymb}
\usepackage{mathtools}
\usepackage{graphicx}
\usepackage{algorithm}
\usepackage{algpseudocode}
\usepackage{bm}
\usepackage{url}
\usepackage{xcolor}
\usepackage{comment}
\usepackage{nomencl} 
\usepackage{array}
\usepackage{cuted}
\usepackage{float}
\usepackage{stfloats}  
\usepackage{tikz}
\usetikzlibrary{arrows.meta, positioning, shapes.geometric}
\usetikzlibrary{calc,shadows.blur}

\newcommand{\vones}{\bm{1}}
\newcommand{\vzeros}{\bm{0}}

\definecolor{harry}{RGB}{0, 128, 192}

\definecolor{delete}{RGB}{240, 0, 0}

\definecolor{Ilyas}{RGB}{216 82 24}

\usepackage[normalem]{ulem}

\newcommand{\define}{\coloneqq}

\begin{document}

\title{Frequency Nadir-Constrained Power System Restoration Planning with Energy Storage}

\author{Xiangyu Zou, Amir Reza Nikzad, Ilyas Farhat, and John W. Simpson-Porco%
\thanks{X.~Zou is with the School of Electrical and Computer Engineering, Georgia Institute of Technology, USA. Email: xy.zou@gatech.edu. A.~R.~Nikzad and J.~W.~Simpson-Porco are with the Department of Electrical and Computer Engineering, University of Toronto, Canada. (Email: amirreza.nikzad@utoronto.ca; jw.simpson@utoronto.ca). I.~Farhat is with EPE Canada Inc., Email: ilyas.farhat@outlook.com.}
\thanks{Research supported by Hydro One Networks Inc. and NSERC ALLRP 597475-2024 ``Planning Restoration of Inverter-Based Grids''.}
}

\maketitle

\begin{abstract}
Power system restoration following blackouts must ensure frequency stability throughout the recovery process. This paper proposes a frequency-constrained mixed-integer linear programming (MILP) framework for black-start restoration planning in transmission systems with synchronous machines and energy storage systems. 
To prevent excessive frequency deviations caused by restorative actions, a frequency nadir prediction method is developed for power systems with energy storage system (ESS) integration and incorporated into a multi-period optimization framework.
The formulation ensures that frequency deviations resulting from restorative actions remain within prescribed safe limits. Furthermore, the presented framework leverages ESSs to enhance frequency security and recovery speed. Case studies on a modified IEEE 9-bus system demonstrate that the computed restoration plan maintains frequency security, as validated through MATLAB and PSS/E simulations, while reducing restoration time through ESS coordination.
\end{abstract}


\section{Introduction} \label{sec1-Introduction}

Power systems face increasing threats from severe weather events and cyber attacks~\cite{Elizabeth2020}, with outages causing up to \$50~billion in annual losses in the United States~\cite{Madeline2023}. Individual large-scale events, such as the 2003 Northeastern blackout, affected 50~million people and resulted in approximately \$6~billion in economic losses~\cite{NERC2004}. Since outage-related damages increase with restoration duration~\cite{Madeline2023}, fast grid recovery strategies are critical. Consequently, Independent System Operators (ISOs) proactively develop restoration plans that include guidelines and predefined actions to support decision-making during grid recovery~\cite{NERC2018}.

Following complete blackouts, black-start units (BSUs) \textemdash{} typically hydro or combustion turbines capable of self-starting \textemdash{} serve as the initial sources of power for non-black-start units (NBSUs) that cannot self-start~\cite{Knight2013}. Black-start restoration proceeds through four phases: sectionalization, generator recovery, load recovery, and synchronization~\cite{Qiu2017}. The grid is first sectionalized into multiple subsystems, each with one BSU, to be restored in parallel~\cite{Qiu2017}. During generator recovery, BSUs prioritize energizing paths toward NBSUs to provide cranking power while picking up critical infrastructure loads, or other loads for power balance and voltage support~\cite{IESO2024}. Early generator recovery increases available capacity and improves system resilience by increasing inertia~\cite{IESO2024}. Once most or all generators have been reconnected, the remaining loads and lines are energized. Finally, the restored subsystems are synchronized to reassemble the complete grid.

Restoration planning involves a fundamental trade-off between speed and reliability. While faster restoration reduces the duration of the outage, it requires more aggressive restorative actions which may generate unacceptable transient disturbances. Each restoration action perturbs the system; for example, load pick-ups introduce power imbalances that cause frequency decline. Frequency issues in particular are exacerbated by the increasing penetration of inverter-based resources (IBRs), which reduce system inertia and increase susceptibility to frequency deviations~\cite{Anderson2003,NREL2020}. Excessive frequency deviations may damage equipment and impede restoration by triggering load-shedding protection, making the frequency nadir—the lowest post-disturbance frequency—a critical security indicator during restoration~\cite{NREL2020}. 

To mitigate frequency risks, operators commonly apply heuristic rules such as limiting load pick-ups to less than 5\% of online generation capacity~\cite{IESO2024}. However, such rules are system-agnostic and do not account for operating conditions, inertia levels, or fast-acting resources. Thus, such rules may be either overly conservative, or overly optimistic, depending on the system state. Moreover, current ISO restoration plans either exclude or do not explicitly consider energy storage systems (ESS)~\cite{MISO2024}, thereby neglecting resources capable of providing rapid power balancing and dynamic frequency support. 

These observations motivate frequency-aware restoration-planning methods that can account for the dynamic effects of planned restoration actions and integrate ESS participation while remaining computationally tractable. The following review places this need in the context of existing restoration-planning and frequency-security methods.

\subsection{Literature Review} \label{LiteratureReview}

Early IEEE reports on restoration can be found in \cite{Adibi1987,Adibi1992}, with a general survey of restoration planning methods in \cite{Yutian2016}; our specific focus will be on \emph{optimization-based} methodologies for power system restoration planning, where the sequence of restoration actions is computed by minimizing an objective function subject to constraints. Reference~\cite{Coffrin2015} develops an ordering problem for line and load recovery coordinating crew dispatch with operator actions, which was extended in~\cite{Yunyun2024} to use mobile energy storage systems. The ordering problem is further extended in~\cite{Van2015} to incorporate rotor angle stability by discretizing governing ODEs as constraints. An algorithm that iteratively refines restoration sequences is presented in~\cite{Noah2022}, achieving near-optimal solutions with reduced computation time. Reference~\cite{Qiu2017} integrates system partitioning with generator and load recovery decisions. Online restoration approaches using real-time measurements include feedback control with optimization-based controllers~\cite{Miller2022}. Deep reinforcement learning has been applied to distribution system restoration~\cite{Mosayeb2023}, achieving fast solution times but focusing on load restoration without explicit frequency stability constraints required for transmission-level black-start. 
Reference~\cite{Papasani2021} studies restoration of industrial facilities, highlighting the need to explicitly consider cold load pickup and motor inrush currents. Recent works also address renewable-dominated grids~\cite{Changming2022}, distributed energy resources~\cite{Braun2018}, and computational challenges through approximate dynamic programming~\cite{Sharma2020}. Reviews emphasize learning from historical blackouts and developing robust methodologies~\cite{Hasini2025}.

Explicit consideration of frequency dynamics in restoration planning remains limited. Reference~\cite{Golshani2019} constrains load pick-up decisions using the frequency-response model of~\cite{Baldick2014}, which relates the frequency nadir to the size of a disturbance through aggregate linear generator-response assumptions. This provides a tractable way to impose frequency-security limits, but does not account for important dynamic effects, including governor--turbine dynamics, saturation limits, and restoration-stage conditions such as online inertia, load damping, and available regulating capacity. 

More broadly, frequency-aware restoration planning faces a modeling--tractability trade-off. Frequency-security studies often assume a fixed network configuration, constant aggregate parameters such as inertia, and single-period or quasi-steady-state operating conditions~\cite{Qingxin2018}. Restoration planning, in contrast, is inherently multi-stage and combinatorial: generator start-ups, line energizations, load pick-ups, and resource-dispatch decisions change the system configuration and dynamic response from one stage to the next. Embedding post-disturbance frequency dynamics into such a mixed-integer planning problem therefore requires closed-form constraints that capture how the frequency nadir depends on discrete restoration decisions and evolving system conditions. Existing approximations either simplify the dynamics enough to lose this dependence or lead to nonlinear constraints that are difficult to incorporate directly into restoration optimization~\cite{Liu2020}.

Taken together, the review highlights a modeling--tractability gap in frequency-aware restoration planning. Existing approaches either rely on tractable but simplified frequency-security rules, or use richer dynamic representations that are difficult to embed directly into large-scale mixed-integer restoration problems. This motivates restoration-planning formulations that retain explicit, stage-dependent frequency-nadir constraints while remaining computationally tractable, and that coordinate fast-acting resources such as ESS to support aggressive restoration actions.

\subsection{Contributions}\label{Contributions}
We develop an optimization-based black-start restoration-planning framework that explicitly accounts for changing inertia and primary frequency dynamics, and enables coordinated ESS support during restoration. A key observation is that, unlike unplanned disturbances, restoration actions are scheduled by the operator. As such, their dynamic impacts can be anticipated and mitigated through coordinated redispatch of both generators and ESS. We exploit this by embedding frequency-nadir predictions directly into a MILP restoration planning formulation, enabling restoration plans that maximize restoration speed subject to frequency security. 

Our preliminary work in~\cite{EPEC2025} sketched the general MILP planning framework, introduced the nadir prediction concept without ESS, and contained a proof-of-concept  case study. This paper expands significantly on \cite{EPEC2025} and provides (i) complete derivations of the MILP planning and nadir prediction frameworks, (ii) incorporation of ESS into the MILP planning and nadir prediction frameworks, and (iii)  case studies in both MATLAB and PSS/E, highlighting how the restoration plans produced by our method lead to frequency-secure black-start restoration sequences. Our results demonstrate that energy storage can be used in surprising and non-intuitive ways to accelerate system restoration through strategic charging and discharging, while ensuring frequency constraints are maintained throughout the restoration sequence.

\subsection{Paper Organization}\label{PaperOrganization} Section~\ref{sec-Freq_const_MILP} develops the MILP formulation for restoration planning. Section~\ref{sec-Freq-Nadir-Estimation} derives the frequency nadir prediction methodology with ESS support. Section~\ref{sec-frequency-cons} integrates frequency constraints into the MILP through an iterative algorithm. Section~\ref{CaseStudy} provides case studies and numerical validation on test systems. Finally, Section~\ref{Conclusion} concludes the paper and discusses future research directions.

\section{MILP for Restoration Planning} \label{sec-Freq_const_MILP}

This section presents the MILP framework for transmission system black-start restoration planning. The model focuses on generator and load recovery within a single subsystem following sectionalization, seeking an optimal action sequence that minimizes restoration time while respecting physical and logical constraints. Loads are modeled as discrete packets representing individual distribution feeders, which constitute the smallest load units that can be independently energized during restoration~\cite{Birchfield2022}. Although MILP models for transmission-level restoration have been developed in other literature~\cite{Qiu2017}, our proposed model is conducive to the integration of frequency limit constraints, as is shown in subsequent sections.

\subsection{Network Logic and Power Flow Constraints}\label{Network-Element-Status}

A restoration plan consists of a sequence of discrete operator actions, such as switching on a network component. Actions may occur every $t_a$ minutes, giving time for transient dynamics to settle between steps. As notation throughout:
\begin{itemize}
\item An unbolded, unitalicized, uppercase symbol such as ${\rm A} \in \mathbb{R}^{M \times N}$ denotes a \emph{constant} matrix; a bolded lowercase symbol such as $\boldsymbol{a} \in \mathbb{R}^{N \times T}$ denotes a \emph{variable} matrix with $N$ rows defined over $T$ time steps.
\item $\boldsymbol{a}[k] \in \mathbb{R}^{N \times 1}$ denotes the $k$th column of $\boldsymbol{a}$, i.e., the values at time step $k \in \{1,\ldots,T\}$;
\item $\boldsymbol{a}^i \in \mathbb{R}^{1 \times T}$ denotes the $i$th row of $\boldsymbol{a}$, i.e., the values of element $i$ over the restoration horizon.
\item $\vones_{M \times N}$ and $\vzeros_{M \times N}$ denote matrices of ones and zeros of size $M \times N$, respectively; $\vones_{N}$ and $\vzeros_{N}$ denote column vectors of length $N$.
\end{itemize}

Binary variables model the on/off status of network elements over $T$ discrete time steps, where $T$ is the total number of time steps during restoration. Let $\boldsymbol{b}_{\rm b} \in \{0,1\}^{B \times T}$, $\boldsymbol{b}_{\rm l} \in \{0,1\}^{L \times T}$, $\boldsymbol{b}_{\rm d} \in \{0,1\}^{D \times T}$,  $\boldsymbol{b}_{\rm g} \in \{0,1\}^{G \times T}$, and $\boldsymbol{b}_{\rm s} \in \{0,1\}^{S \times T}$ denote the status matrices for $B$ buses, $L$ lines, $D$ loads, $G$ generators, and $S$ ESSs, respectively. The stacked status matrix $\boldsymbol{b} = \mathrm{col}(\boldsymbol{b}_{\rm b}, \boldsymbol{b}_{\rm l}, \boldsymbol{b}_{\rm d}, \boldsymbol{b}_{\rm g}, \boldsymbol{b}_{\rm s})$ encodes the complete restoration plan. 
The initial status at time $k=0$ is specified by the vector
$\boldsymbol{b}_0 = \mathrm{col}(\boldsymbol{b}_{\rm b,0}, \boldsymbol{b}_{\rm l,0}, \boldsymbol{b}_{\rm d,0}, \boldsymbol{b}_{\rm g,0}, \boldsymbol{b}_{\rm s,0})=\boldsymbol{b}[0]$. For black-start scenarios, each subsystem contains a single BSU following sectionalization. Furthermore, all network elements are offline except for the BSU and its connected bus.

Network elements must obey two key activation rules for all $k \in \{1,\ldots,T\}$. First, once any element is restored, it must remain connected, written as
\begin{equation}
    \boldsymbol{b}[k] - \boldsymbol{b}[k-1] \geq \vzeros,
    \label{eq-milp-stayon}
\end{equation}
where inequalities are element-wise. Second, at most one element of each type may be activated per time step
\begin{equation}
    \vones_{N_x}^{\sf T}\bigl(\boldsymbol{b}_{x}[k] - \boldsymbol{b}_{x}[k-1]\bigr) \leq 1,
    \qquad x \in \{\rm b,l,d,g,s\}
    \label{eq-milp-eleperstep}
\end{equation}
where $N_x$ is the number of elements of type $x$ and the notation $v^{\sf T}$ denotes the transpose of $v$.

Element network locations are encoded by the element-to-bus adjacency matrices
${\rm A_{l}} \in \{0,1\}^{B \times L}$,
${\rm A_{d}} \in \{0,1\}^{B \times D}$,
${\rm A_{g}} \in \{0,1\}^{B \times G}$, and
${\rm A_{s}} \in \{0,1\}^{B \times S}$,
where $[\rm {A}_{\bullet}]_{b,a} = 1$ if element $a$ is connected to bus $b$, and 0 otherwise. The bus-line incidence matrix ${\rm A} \in \mathbb \{-1,0,1\}^{B \times L}$ defines the network topology and positive line flow orientations~\cite{Bullo2018}.

The statuses of network elements are linked by logical constraints to ensure valid restoration sequences. Energizing a line implies energizing both of its terminal buses \eqref{eq-milp-ltob1}; buses are necessarily offline if no connected lines are energized, except for initially active buses \eqref{eq-milp-ltob2}; and lines may be energized only if at least one terminal bus is already active \eqref{eq-milp-ltob3}:
\begin{subequations}\label{eq-milp-ltob}
\begin{align}
    {{\rm A}_{\rm l}^{(i)}} \boldsymbol{b}_{\rm l}^i &\leq \boldsymbol{b}_{\rm b},
        &&  i \in \{1, \dots, L\}, \label{eq-milp-ltob1} \\
    {\rm{A}_{l}} \boldsymbol{b}_{\rm l} &\geq \boldsymbol{b}_{\rm b} - \boldsymbol{b}_{\rm b,0}\vones_{T}^\top,
        && \label{eq-milp-ltob2} \\
    {\rm{A}_{l}^{\top}} \boldsymbol{b}_{\rm b}[k-1] &\geq \boldsymbol{b}_{\rm l}[k],
        &&  k \in \{1, \ldots, T\}, \label{eq-milp-ltob3}
\end{align}
\end{subequations}
where $\rm{A}_{l}^{(i)}$ denotes the $i$th column of $\rm{A}_{l}$.

Loads and ESSs may be energized only if their associated bus is active, as expressed in \eqref{eq-milp-dgtob-d} and \eqref{eq-milp-dgtob-s}; generators additionally require their bus to be active for at least one time step prior to start-up, as expressed in \eqref{eq-milp-dgtob-g}:
\begin{subequations}\label{eq-milp-dgtob}
\begin{align}
    {\rm{A}_{d}} \boldsymbol{b}_{\rm d}[k] &\leq M \boldsymbol{b}_{\rm b}[k] \label{eq-milp-dgtob-d}, \\
    {\rm{A}_{s}} \boldsymbol{b}_{\rm s}[k] &\leq M \boldsymbol{b}_{\rm b}[k] \label{eq-milp-dgtob-s}, \\
    {\rm{A}_{g}} \boldsymbol{b}_{\rm g}[k+1] &\leq M \boldsymbol{b}_{\rm b}[k] \label{eq-milp-dgtob-g},
\end{align}
\end{subequations}
which must hold for all $k \in \{0,\ldots,T-1\}$. Here, and throughout, $M$ is a sufficiently large constant used to enforce these conditional constraints via the big-M method.\footnote{The big-M method is a standard MILP technique for enforcing conditional constraints. The value of $M$ should be chosen as a tight upper bound, since excessively large values may cause numerical ill-conditioning.}

The active power balance at every bus and every time step is enforced via 
\begin{equation}\label{eq-milp-DCPF}
    {\rm{A}_{g}} {\boldsymbol{P}_{\rm g}}
    - {\rm{A}_{d}}\mathrm{diag}(\boldsymbol{P}_{\rm d}) \boldsymbol{b}_{\rm d}
    - {\rm{A}} \boldsymbol{P}_{\rm l}
    + {\rm{A}_{s}} \boldsymbol{P}_{\rm s}
    = \vzeros_{B \times T},
\end{equation}
where $\boldsymbol{P}_{\rm d} \in \mathbb{R}^{D}$ is the fixed vector of load magnitudes, $\boldsymbol{P}_{\rm l} \in \mathbb{R}^{L \times T}$ contains line flows, $\boldsymbol{P}_{\rm g} \in \mathbb{R}^{G \times T}$ contains generator outputs, and $\boldsymbol{P}_{\rm s} \in \mathbb{R}^{S \times T}$ contains net ESS power injections.

We adopt the DC power flow (DCPF) approximation to define active power flows for energized lines ~\cite{DCPF}, and flows across de-energized lines are set to zero. 
Defining the complement status matrix $\tilde{\boldsymbol{b}}_{x} \triangleq \vones_{N_x \times T} - \boldsymbol{b}_{x}$, these two conditions can be expressed as
\begin{subequations}\label{eq-milp-PL}
\begin{align}
    -M\tilde{\boldsymbol{b}}_{\rm l} \leq \boldsymbol{P}_{\rm l} 
        - {\rm X}^{-1}\mathrm{A}^{\top}\boldsymbol{\theta}
        &\leq M\tilde{\boldsymbol{b}}_{\rm l}  \label{eq-milp-PLa},\\
    -M\boldsymbol{b}_{\rm l} \leq \boldsymbol{P}_{\rm l}
        &\leq M\boldsymbol{b}_{\rm l},\label{eq-milp-PLb}
\end{align}
\end{subequations}
where $\mathrm{X}\in \mathbb{R}^{L\times L}$ is the diagonal matrix of line reactances and $\boldsymbol{\theta}\in \mathbb{R}^{B\times T}$ is the matrix of bus voltage phase angles. Finally,
\begin{equation}
    -M\boldsymbol{b}_{\rm b} \leq \boldsymbol{\theta} \leq M\boldsymbol{b}_{\rm b}
    \label{eq-milp-thetacon}
\end{equation}
ensures that phase angles at de-energized buses and at the BSU reference bus are set to zero.

\subsection{NBSU Start-up Process}\label{NBSU-Start}
\begin{figure}[h!]
\centering
\includegraphics[width=0.75\linewidth]{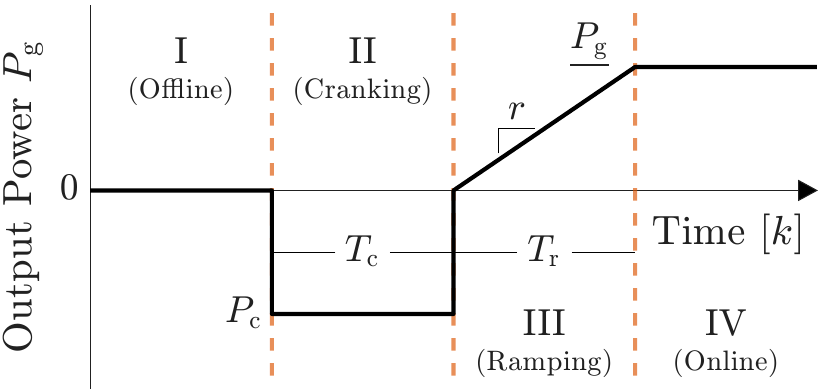}
\caption{Four phases of NBSU start-up.}
\label{fig-nbsu-phases}
\end{figure}

The start-up of NBSUs proceeds through four phases~\cite{Qiu2017}: offline, cranking, ramping, and online, as illustrated in Fig.~\ref{fig-nbsu-phases}. During cranking (duration $T_{\rm c}^i$ steps), generator $i$ consumes cranking power $P_{\rm c}^i$. During ramping (duration $T_{\rm r}^i$ steps), its output increases linearly from zero to the minimum output $\underline{P}_{\rm g}^i$ at a ramp rate $r^i$ (in units p.u. power per step). Once online, the generator operates within power limits $[\underline{P}_{\rm g}^i, \overline{P}_{\rm g}^i]$ and ramp-rate limits $[-r^i, r^i]$.

To capture the start-up phase of each generator in the MILP framework, auxiliary binary variables $\boldsymbol{b}_{\rm gc}, \boldsymbol{b}_{\rm gr}, \boldsymbol{b}_{\rm go} \in \{0,1\}^{G \times T}$ are introduced, where $\boldsymbol{b}_{\rm gc}^i[k] = 1$ if generator $i$ is in the cranking phase at step $k$, $\boldsymbol{b}_{\rm gr}^i[k] = 1$ if it is ramping, and $\boldsymbol{b}_{\rm go}^i[k] = 1$ if it is online; all three are zero when the generator is offline. Since the cranking and ramping durations are known, these variables are fully determined by the time step at which the generator is switched on and can be expressed in terms of $\boldsymbol{b}_{\rm g}$ for all $k \in \{1,\ldots,T\}$ and $i \in \{1,\ldots,G\}$ via
\begin{subequations}\label{eq-milp-aux}
\begin{align}
    \boldsymbol{b}_{\rm gc}^i[k] &= \boldsymbol{b}_{\rm g}^i[k] 
        - \boldsymbol{b}_{\rm g}^i[k - T_{\rm c}^i] \label{eq-milp-bgc},\\
    \boldsymbol{b}_{\rm gr}^i[k] &= \boldsymbol{b}_{\rm g}^i[k - T_{\rm c}^i] 
        - \boldsymbol{b}_{\rm g}^i[k - T_{\rm c}^i - T_{\rm r}^i] \label{eq-milp-bgr},\\
    \boldsymbol{b}_{\rm go}^i[k] &= \boldsymbol{b}_{\rm g}^i[k] 
        - \boldsymbol{b}_{\rm gc}^i[k] - \boldsymbol{b}_{\rm gr}^i[k]. \label{eq-milp-bgo}
\end{align}
\end{subequations}
For time indices $k \leq 0$ preceding restoration, all NBSUs are offline, i.e., $\boldsymbol{b}_{\rm g}^i[k] = 0$. All per-generator scalar parameters are collected into vectors $P_{\rm c}, r, \underline{P}_{\rm g},
\overline{P}_{\rm g} \in \mathbb{R}^G$.

To compactly express the phase-dependent output constraints in Fig.~\ref{fig-nbsu-phases}, define the transient mask
$\boldsymbol{\Phi} \triangleq \boldsymbol{b}_{\rm gc} + \boldsymbol{b}_{\rm gr}$,
which equals one whenever a generator is in a transient phase (cranking or ramping),
and let $\Delta\boldsymbol{P}_{\rm g}[k] = \boldsymbol{P}_{\rm g}[k] -
\boldsymbol{P}_{\rm g}[k-1]$ denote the change in generator power outputs at step $k$. The behavior of Fig.~\ref{fig-nbsu-phases} is enforced via 
\begin{subequations}\label{eq-milp-nbsu-cro}
\begin{align} 
    -M\tilde{\boldsymbol{b}}_{\rm gc} \leq \boldsymbol{P}_{\rm g} 
        + \boldsymbol{P}_{\rm c} \vones_{T}^\top
        &\leq M\tilde{\boldsymbol{b}}_{\rm gc},
        \label{eq-milp-nbsu-croa}\\
    -M\tilde{\boldsymbol{b}}_{\rm gr} \leq \boldsymbol{P}_{\rm g} 
        - \boldsymbol{P}_{\rm r}
        &\leq M\tilde{\boldsymbol{b}}_{\rm gr},
        \label{eq-milp-nbsu-crob}\\
    \mathrm{diag}(\underline{\boldsymbol{P}}_{\rm g}) \boldsymbol{b}_{\rm go} 
        - M\boldsymbol{\Phi} \leq \boldsymbol{P}_{\rm g}
        &\leq \mathrm{diag}(\overline{\boldsymbol{P}}_{\rm g}) \boldsymbol{b}_{\rm go} 
        + M\boldsymbol{\Phi},
        \label{eq-milp-nbsu-croc} \\
        -r - M\tilde{\boldsymbol{b}}_{\rm go}[k] \leq \Delta\boldsymbol{P}_{\rm g}[k] &\leq r + M\tilde{\boldsymbol{b}}_{\rm go}[k].
        \label{eq-milp-ramp-after}
\end{align}
\end{subequations}
Here, the constraint \eqref{eq-milp-nbsu-croa} enforces the consumption of cranking power during phase II, \eqref{eq-milp-nbsu-crob} ties the output to an auxiliary variable $\boldsymbol{P}_{\rm r}\in \mathbb{R}^{G\times T}$ that describes the linear ramping output during phase III, \eqref{eq-milp-nbsu-croc} enforces the output limits during phase IV, and \eqref{eq-milp-ramp-after} enforces the ramping limits during phase IV.

The ramp reference $\boldsymbol{P}_{\rm r}$ is defined by the constraints 
\begin{subequations}\label{eq-milp-ramping}
\begin{align}
    -M\boldsymbol{b}_{\rm gr} \leq \boldsymbol{P}_{\rm r} 
        + \tfrac{1}{2}r \vones_{T}^\top
        &\leq M\boldsymbol{b}_{\rm gr},
        \label{eq-milp-NBSU-pra} \\
    -M\tilde{\boldsymbol{b}}_{\rm gr}[k] \leq \Delta\boldsymbol{P}_{\rm r}[k] 
        - r
        &\leq M\tilde{\boldsymbol{b}}_{\rm gr}[k],
        \label{eq-milp-NBSU-prb}
\end{align}
\end{subequations}
to begin increasing at the ramp rate $r$ at the onset of each generator's ramping phase. It is held at a constant value during other phases; this constant value is unused since constraint \eqref{eq-milp-nbsu-crob} is slack. See~\cite{HarryThesis} for further modeling details.

\subsection{Energy Storage System Modeling}\label{ESS-Modelling}
\begin{figure}[h!]
\centering
\begin{tikzpicture}[
    node distance = 1.2cm and 1.8cm,
    block/.style  = {draw, rectangle, minimum width=1.6cm, minimum height=1.0cm,
                     align=center, font=\small},
    darr/.style   = {{Stealth[length=3pt]}-{Stealth[length=3pt]}, thick},
]

\draw[thick, {Stealth[length=4pt]}-] (-1.5, 0) -- (-0.02, 0);
\node[below, font=\footnotesize] at (-0.75, 0) {Grid};
\draw[line width=2pt] (0,-0.4) -- (0,0.4);
\node[above, font=\scriptsize] at (0,0.45) {Bus $b$};

\node[block, right=1.6cm of {(0,0)}] (conv) {Converter\\$\eta_{\rm con}$};
\node[block, right=1.8cm of conv]    (batt) {Battery\\$\eta_{\rm s}$};

\draw[darr] (0,0) -- node[above, font=\small]{${P}_{\rm s}$}
                     node[below, font=\footnotesize]{AC}
            (conv.west);

\draw[darr] (conv.east) --
    node[above, font=\small]{${P}_{\rm s}^{\rm in},\,{P}_{\rm s}^{\rm out}$}
    node[below, font=\footnotesize]{DC}
    (batt.west);

\node[below=0.05cm of batt, font=\footnotesize] {${E}_{\rm s}$};

\end{tikzpicture}
\caption{ESS grid interconnection.}
\label{ESS-Converter}
\end{figure}
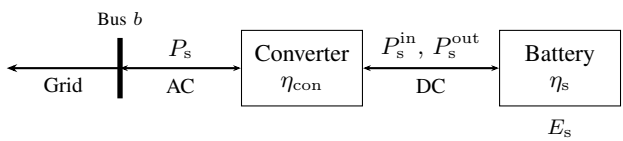

ESS devices can provide operational flexibility during restoration by storing excess generation when supply exceeds demand and supplying power during load recovery when generator ramping capability is insufficient to support large load pick-ups.
We consider a battery ESS model in which a storage device exchanges power with the grid through a DC-AC converter, as illustrated in Fig.~\ref{ESS-Converter}. The efficiencies of the converter and storage elements for each ESS are given by the vectors $\eta_{\rm con}, \eta_{\rm s} \in [0,1]^{S}$, respectively.

The state of charge (energy stored) is denoted by $\boldsymbol{E}_{\rm s} \in \mathbb{R}^{S \times T}$, with initial value $E_{\rm s,0} \in \mathbb{R}^S$, and is bounded by the maximum storage capacity $\overline{E}_{\rm s} \in \mathbb{R}^S$, i.e., 
\begin{equation} \label{eq-soc-bound}
    \vzeros_{S} \leq \boldsymbol{E}_{\rm s}[k]
    \leq \overline{E}_{\rm s},
    \qquad  k \in \{1,\ldots,T\}.
\end{equation}

The power flow to and from the battery is decomposed into nonnegative 
charging and discharging components
$\boldsymbol{P}_{\rm s}^{\rm in}, \boldsymbol{P}_{\rm s}^{\rm out} \in 
\mathbb{R}^{S \times T}$, where $\boldsymbol{P}_{\rm s}^{\rm in}[k]
, \boldsymbol{P}_{\rm s}^{\rm out}[k] \geq \vzeros_{S}$.
Accounting for losses, the AC-side net power injected into the grid is
\begin{equation}\label{eq-ps}
    \boldsymbol{P}_{\rm s}
    = \mathrm{diag}(\eta_{\rm con}) \boldsymbol{P}_{\rm s}^{\rm out}
    - \mathrm{diag}(\eta_{\rm con}^{-1}) \boldsymbol{P}_{\rm s}^{\rm in},
\end{equation}
where $\eta_{\rm con}^{-1}$ denotes the element-wise inverse of the converter efficiency vector. Binary variables $\boldsymbol{b}_{\rm s}^{\rm in}, \boldsymbol{b}_{\rm s}^{\rm out} 
\in \{0,1\}^{S \times T}$ indicate active charging and discharging states. Rated 
power limits $\overline{P}_{\rm s} \in \mathbb{R}^S$ are enforced as
\begin{subequations} \label{eq-ess-psto}
\begin{align}
    \vzeros_{S} \leq \boldsymbol{P}_{\rm s}^{\rm in}[k]
        &\leq \overline{P}_{\rm s} \odot \boldsymbol{b}_{\rm s}^{\rm in}[k], \\
    \vzeros_{S} \leq \boldsymbol{P}_{\rm s}^{\rm out}[k]
        &\leq \overline{P}_{\rm s} \odot \boldsymbol{b}_{\rm s}^{\rm out}[k],
\end{align}
\end{subequations}
for all $k \in \{1,\ldots,T\}$, where $\odot$ denotes element-wise multiplication. Simultaneous charging and discharging is prohibited,
and ESS operation requires bus energization, written as
\begin{equation}
    \boldsymbol{b}_{\rm s}^{\rm in}[k] + \boldsymbol{b}_{\rm s}^{\rm out}[k]
    \leq \boldsymbol{b}_{\rm s}[k].
    \label{eq-ess-bin}
\end{equation}
The ESS state of charge evolves according to
\begin{equation} \label{eq-soc-update-updated}
    \boldsymbol{E}_{\rm s}[k+1]
    = \boldsymbol{E}_{\rm s}[k]
    + \frac{t_a}{60}\Bigl(
        \eta_{\rm s} \odot \boldsymbol{P}_{\rm s}^{\rm in}[k]
        - \eta_{\rm s}^{-1} \odot \boldsymbol{P}_{\rm s}^{\rm out}[k]
    \Bigr),
\end{equation}
Storage efficiency is accounted for in \eqref{eq-soc-update-updated}, while the converter efficiency is accounted for in \eqref{eq-ps}, capturing losses at different points in the ESS chain.
Inter-temporal power variation is constrained by the ESS ramp rate 
$r_{\rm s} \in \mathbb{R}^S$ as
\begin{equation} \label{eq-psto-rate}
    -r_{\rm s} \leq
    \boldsymbol{P}_{\rm s}[k+1] - \boldsymbol{P}_{\rm s}[k]
    \leq r_{\rm s},
    \qquad  k \in \{1,\ldots,T-1\}.
\end{equation}

The complete model allows excess power to be deposited at one step and withdrawn at another step with some loss.

\subsection{Objective Function}\label{ObjectiveFunction}
The objective of restoration is the rapid recovery all network elements, which can be achieved by maximizing the statuses of all elements across the time period $T$ written as
\begin{equation} \label{eq-milp-obj}
    J = \sum_{k=1}^{T} \Bigl(
        \boldsymbol{b}_{\rm g}[k]^{\sf T} w_{\rm g}
        + \boldsymbol{b}_{\rm d}[k]^{\sf T} w_{\rm d}
        + \boldsymbol{b}_{\rm l}[k]^{\sf T} w_{\rm l}
        + \boldsymbol{b}_{\rm s}[k]^{\sf T} w_{\rm s}
    \Bigr).
\end{equation}
The weight vectors $w_{\rm g} \in \mathbb{R}^G$, $w_{\rm d} \in \mathbb{R}^D$, 
$w_{\rm l} \in \mathbb{R}^L$, and $w_{\rm s} \in \mathbb{R}^S$, represent the relative priority of
generators, loads, lines, and ESSs, respectively, where greater weights 
indicate higher priority. All weights are dimensionless and can be assigned to reflect operator preferences or system-specific 
requirements. For example, the elements of $w_{\rm g}$ can be set to prioritize larger generators that provide more power capacity and inertia once connected. The elements of $w_{\rm d}$ may depend on the relative sizes of each load, and whether they represent critical infrastructure. In this work, the weights are assigned such that the order of magnitudes follow $w_{\rm g} \gg w_{\rm d} \gg 
w_{\rm l},w_{\rm s}$ to encourage early generator recovery, which increases available 
generation capacity and system inertia and aligns with IESO restoration guidelines~\cite{IESO2024}. Lines and ESSs are given relatively smaller weights as they will naturally be restored to facilitate the primary goal of generator and load recovery.

The complete (frequency-agnostic) restoration problem is 
\begin{equation}\label{Eq:TrueMILP}
\text{maximize}\,\, \eqref{eq-milp-obj} \quad \text{subject to} \quad  \eqref{eq-milp-stayon}-\eqref{eq-psto-rate}.
\end{equation}

\section{Frequency Nadir Estimation with ESS Support}\label{sec-Freq-Nadir-Estimation}
This section derives analytical frequency nadir predictions for step electrical disturbances, accounting for synchronous generator primary frequency response (PFR) and ESS contributions. These predictions yield constraints on the maximum allowable electrical disturbance at each restoration step, which are integrated into the MILP formulation of Section~\ref{sec-Freq_const_MILP} to ensure frequency-secure restoration plans. The analysis focuses on power-shortage scenarios—caused by load pick-ups, generator cranking, and ESS setpoint changes—where frequency declines following a disturbance; an analogous approach can be taken for power-surplus scenarios.

\subsection{Frequency Dynamics Modeling}\label{subsec:frequency}
When a power imbalance occurs during restoration, system frequency undergoes a transient decline. The frequency nadir\textemdash the lowest point of this decline\textemdash is typically reached within 1--10~s following the disturbance, after which primary frequency response (PFR) from generators arrests the decline. We seek to ensure the nadir is within a safe, specified limit. 

\begin{figure}[h!]
    \centering
    \includegraphics[width=0.9\linewidth]{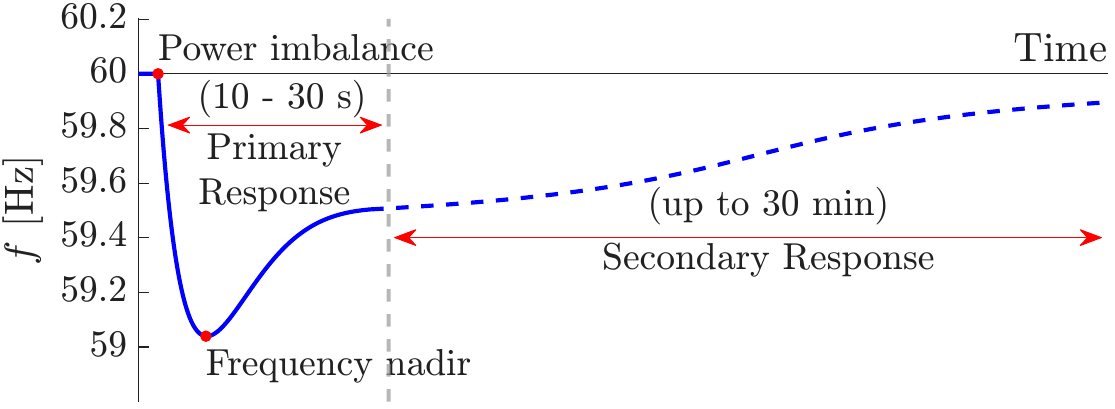}
    \caption{Response of frequency response following a disturbance.}
    \label{fig:freq-phases}
\end{figure}

To model the transient frequency bahevior, we adopt the average system frequency (ASF) model~\cite{Chan1972}, which is well-suited for the seconds timescale. Rather than modeling local frequencies at each generator bus via differential-algebraic equations, the ASF model describes a unified center-of-inertia frequency deviation $\Delta\omega$, which evolves as
\begin{equation} \label{ASF-multi}
    2H_{\rm sys}\Delta\dot{\omega}
    = \sum_{i \in \mathcal{G}_{\rm pfr}} \nolimits \alpha^i \Delta P_{\rm m}^i - \Delta P_{\rm e},
\end{equation}
where $\Delta P_{\rm e}$ is the net electrical disturbance, $\Delta P_{\rm m}^i$ is the mechanical power response of generator $i \in \mathcal{G}_{\rm pfr}$ (the set of PFR-providing generators), and $\alpha^i = S_{\rm g}^i / S_{\rm sys}$ converts each generator's output from its own base $S_{\rm g}^i$ to the system base $S_{\rm sys}$. The system inertia constant is
\begin{equation} \label{Hsys}
    H_{\rm sys} = \tfrac{1}{S_{\rm sys}}\sum_{i \in \mathcal{G}_{\rm on}}\nolimits H^i S_{\rm g}^i,
\end{equation}
where $\mathcal{G}_{\rm on}$ is the set of synchronized generators, and $H^i$ is the inertia constant of unit $i$. 
Given a frequency nadir limit $\Delta\omega_{\rm lim} < 0$, the objective is to determine the maximum electrical disturbance $\Delta P_{\rm e}$ such that $\Delta\omega(t) \geq \Delta\omega_{\rm lim}$ for all $t \geq 0$. This can be ensured by instead bounding the frequency nadir $\Delta\omega_{\rm nad}$; see Fig~\ref{fig:freq-phases}.

\subsection{Governor-Turbine and ESS Dynamics}\label{IEEEG1-Governer-Turbine}

A load pick-up introduces a net power shortage which causes a frequency decline. In response, the governors of synchronous machines participating in PFR increase fuel infeed to their turbines~\cite{Chow2020}. This increases their mechanical power outputs $\Delta P_{\rm m}^i$ in \eqref{ASF-multi}, and stabilizes the system. Simultaneously, ESSs can adjust their power setpoints. We describe the models used to represent these processes.

\subsubsection{IEEEG1 Governor-Turbine Model}
\begin{figure}[h!]
    \centering

\begin{center}
\resizebox{\columnwidth}{!}{%
\begin{tikzpicture}[auto, every node/.style={scale=0.72}]

    \tikzstyle{sum} = [draw, fill=none, circle, node distance=1cm]
    \tikzstyle{input} = [coordinate]
    \tikzstyle{output} = [coordinate]
    \tikzstyle{anch} = [coordinate]
    \tikzstyle{block} = [draw, fill=white, rectangle,
        minimum height=3em, minimum width=4em]
    \tikzstyle{wideblock} = [draw, fill=white, rectangle,
        minimum height=3em, minimum width=5.8em]
    \tikzstyle{smallblock} = [draw, fill=white, rectangle,
        minimum height=3em, minimum width=1.45em]
    \tikzstyle{satblock} = [draw, fill=white, rectangle,
        minimum height=3.2em, minimum width=1.15em,
        inner xsep=0pt, inner ysep=0pt]
    \tikzstyle{gain} = [draw, fill=white, rectangle,
        minimum height=2.1em, minimum width=2.4em]
    \tikzstyle{arr} = [thick, -{Latex[length=1.45mm,width=1.15mm]}]

    \node[input] (in) at (0,0) {};
    \node[wideblock] (droop) at (1.08,0)
        {\!$\displaystyle \frac{K(1+sT_2)}{1+sT_1}$\!};

    \node[sum, minimum size=1.08em, inner sep=0pt] (sum0) at (2.24,0) {};
    \node[smallblock] (T3) at (2.88,0)
        {$\displaystyle \frac{1}{T_3}$};

    \node[satblock] (sat1) at (3.48,0)
        {\tikz[scale=0.115, baseline=-0.5ex]{
            \draw[thick] (-1.25,-1.05) -- (-0.35,-1.05) -- (0.35,1.05) -- (1.25,1.05);
        }};
    \node[smallblock] (int1) at (4.02,0)
        {$\displaystyle \frac{1}{s}$};
    \node[satblock] (sat2) at (4.69,0)
        {\tikz[scale=0.115, baseline=-0.5ex]{
            \draw[thick] (-1.25,-1.05) -- (-0.35,-1.05) -- (0.35,1.05) -- (1.25,1.05);
        }};

    \node[anchor=south] at (sat1.north) {$U_o$};
    \node[anchor=north] at (sat1.south) {$U_c$};
    \node[anchor=south] at (sat2.north) {$\overline{P}_g$};
    \node[anchor=north] at (sat2.south) {$\underline{P}_g$};

    \node[block] (T4) at (5.87,0)
        {$\displaystyle \frac{1}{1+sT_4}$};
    \node[block] (T5) at (7.12,0)
        {$\displaystyle \frac{1}{1+sT_5}$};
    \node[block] (T6) at (8.37,0)
        {$\displaystyle \frac{1}{1+sT_6}$};
    \node[block] (T7) at (9.62,0)
        {$\displaystyle \frac{1}{1+sT_7}$};

    \node[gain] (K1) at (6.50,0.82) {$K_1$};
    \node[gain] (K3) at (7.75,0.82) {$K_3$};
    \node[gain] (K5) at (9.00,0.82) {$K_5$};
    \node[gain] (K7) at (10.27,0.82) {$K_7$};

    \node[sum, minimum size=1.08em, inner sep=0pt] (s1) at (7.75,1.62) {};
    \node[sum, minimum size=1.08em, inner sep=0pt] (s2) at (9.00,1.62) {};
    \node[sum, minimum size=1.08em, inner sep=0pt] (s3) at (10.27,1.62) {};

    \draw[red, very thick, dashed]
        (5.17,-0.62) rectangle (11.15,2.05);

    \node[red, anchor=west] at (5.24,1.78) {\large $G_T(s)$};

    \draw[arr] (in) -- node[above, xshift=-0.1cm] {$\Delta\omega$} (droop);
    \draw[arr] (droop) -- node[pos=0.94, above] {$-$} (sum0);

    \draw[arr] (2.24,1.15) -- node[pos=0.10, left] {$\Delta P_{\rm ref}$}
        node[pos=0.88, right] {$+$} (sum0);

    \draw[arr] (sum0) -- (T3);
    \draw[arr] (T3) -- (sat1);
    \draw[arr] (sat1) -- (int1);
    \draw[arr] (int1) -- (sat2);
    \draw[arr] (sat2) -- (T4);

    \node[red] at (3.44,1.08) {\large SAT1};
    \node[red] at (4.65,1.08) {\large SAT2};

    \coordinate (pgtap) at ($(int1.east)!0.45!(sat2.west)$);

    \coordinate (pgfbdown) at ($(pgtap)+(0,-0.90)$);
    \coordinate (pgfbleft) at (2.24,-0.90);
    \draw[thick] (pgtap) -- (pgfbdown);
    \draw[thick] (pgfbdown) -- (pgfbleft);
    \draw[arr] (pgfbleft) -- node[pos=0.88, right] {$-$} (sum0);

    \draw[arr] (T4) -- (T5);
    \draw[arr] (T5) -- (T6);
    \draw[arr] (T6) -- (T7);

    \coordinate (tap1) at ($(T4.east)!0.5!(T5.west)$);
    \coordinate (tap3) at ($(T5.east)!0.5!(T6.west)$);
    \coordinate (tap5) at ($(T6.east)!0.5!(T7.west)$);

    \draw[arr] (tap1) -- (K1.south);
    \draw[arr] (tap3) -- (K3.south);
    \draw[arr] (tap5) -- (K5.south);

    \draw[arr] (T7.east) -- (10.27,0) -- (K7.south);

    \draw[arr] (K1.north) -- (6.50,1.62) -- node[pos=0.88, above] {$+$} (s1);

    \draw[arr] (K3.north) -- node[pos=0.92, right] {$+$} (s1.south);
    \draw[arr] (K5.north) -- node[pos=0.92, right] {$+$} (s2.south);
    \draw[arr] (K7.north) -- node[pos=0.92, right] {$+$} (s3.south);

    \draw[arr] (s1) -- node[pos=0.88, above] {$+$} (s2);
    \draw[arr] (s2) -- node[pos=0.88, above] {$+$} (s3);
    \draw[arr] (s3) -- node[above] {$\Delta P_m$} (11.00,1.62);

\end{tikzpicture}%
}
\end{center}
    \caption{IEEEG1 turbine-governor diagram.}
    \label{IEEEG1-Figure}
\end{figure}
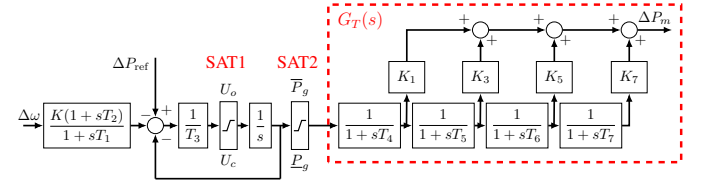

We adopt the IEEEG1 governor-turbine model for steam turbines, shown in Fig.~\ref{IEEEG1-Figure}. The model consists of a governor that detects frequency deviation and sends a control signal to the turbine~\cite{Neplan2015}. The droop constant $K$ determines the proportional relationship between frequency deviation and power increase. A saturation block SAT1 limits the rate at which steam valves can be adjusted, representing the generator ramp rate. A second saturation block SAT2 limits the valve position. The turbine stages are modeled as cascading low-pass filters with time constants $T_4,\ldots,T_7 > 0$. All parameters and signals are in per-unit on each generator's base.

\subsubsection{ESS Dynamic Model}
ESSs are modeled as dispatchable resources with first-order dynamics. Unlike synchronous generators, we assume ESSs do not participate in PFR, but that their setpoints may be adjusted with each restoration action. Each ESS receives a power setpoint $\Delta P_{\rm s}^{{\rm ref},i}$ (positive for discharging) and adjusts its power $\Delta P_{\rm s}^i$ with time constant $\tau^i$, as represented by the transfer function.
\begin{equation}\label{eq-ess-tf}
    G_{\rm s}^{i}(s) = \frac{1}{1+s\tau^{i}}
\end{equation}
This model captures a fast but non-instantaneous ESS response, balancing accuracy and simplicity.

\subsubsection{Closed-Loop System with PFR and ESS}
The closed-loop feedback system with multiple synchronous generators and ESS units is shown in Fig.~\ref{fig-closed-loop}. It is assumed that transients from previous restoration actions have settled prior to the next action, so all signals represent deviations from the prevailing steady state.
\begin{figure}[h!]
    \centering

\begin{center}
\resizebox{0.8\columnwidth}{!}{%
\begin{tikzpicture}[auto, every node/.style={scale=0.62}]

    \tikzstyle{sum} = [draw, fill=none, circle, node distance=1cm]
    \tikzstyle{input} = [coordinate]
    \tikzstyle{output} = [coordinate]
    \tikzstyle{anch} = [coordinate]
    \tikzstyle{block} = [draw, fill=white, rectangle,
        minimum height=3em, minimum width=4em]
    \tikzstyle{wideblock} = [draw, fill=white, rectangle,
        minimum height=3em, minimum width=5.8em]
    \tikzstyle{smallblock} = [draw, fill=white, rectangle,
        minimum height=3em, minimum width=1.45em]
    \tikzstyle{satblock} = [draw, fill=white, rectangle,
        minimum height=3.2em, minimum width=1.15em,
        inner xsep=0pt, inner ysep=0pt]
    \tikzstyle{gain} = [draw, fill=white, rectangle,
        minimum height=2.1em, minimum width=2.4em]
    \tikzstyle{arr} = [thick, -{Latex[length=1.45mm,width=1.15mm]}]

    \tikzstyle{swingblock} = [draw, fill=white, rectangle,
        minimum height=3em, minimum width=4.65em]
    \tikzstyle{redblock} = [draw=red, fill=white, rectangle,
        minimum height=2.15em, minimum width=3.45em,
        inner xsep=1pt, inner ysep=0.5pt]

    \node[input] (in1) at (0,1.05) {};
    \node[input] (inG) at (0,0.00) {};

    \node[block] (lp1) at (1.05,1.05)
        {$\displaystyle \frac{1}{1+s\tau}$};
    \node[block] (lpG) at (1.05,0.00)
        {$\displaystyle \frac{1}{1+s\tau}$};

    \node[sum, minimum size=0.88em, inner sep=0pt] (sess) at (2.02,1.05) {};
    \node[sum, minimum size=0.88em, inner sep=0pt] (smain) at (2.70,1.05) {};

    \node[swingblock] (swing) at (4.04,1.05)
        {$\displaystyle \frac{1}{2H_{\rm sys}s}$};

    \node[output] (out) at (5.78,1.05) {};

    \node[gain] (a1) at (3.50,0.34) {$\alpha_1$};
    \node[gain] (aG) at (3.50,-0.62) {$\alpha_G$};

    \node[sum, minimum size=0.88em, inner sep=0pt] (sgen) at (2.70,0.34) {};

    \node[redblock] (ieee1) at (4.96,0.34) {\scriptsize\color{red} IEEEG1$^{1}$};
    \node[redblock] (ieeeG) at (4.96,-0.62) {\scriptsize\color{red} IEEEG1$^{G}$};

    \node at (1.05,0.60) {$\vdots$};
    \node at (2.70,-0.14) {$\vdots$};
    \node at (3.50,-0.14) {$\vdots$};
    \node at (4.96,-0.14) {$\vdots$};

    \draw[arr] (in1) -- node[pos=0.22, above] {$\Delta P^{\rm ref}_{\rm ess,1}$} (lp1);
    \draw[arr] (inG) -- node[pos=0.22, above] {$\Delta P^{\rm ref}_{\rm ess,E}$} (lpG);

    \draw[arr] (lp1) -- node[pos=0.76, above] {$+$} (sess);

    \draw[arr] (lpG.east) -- (sess |- lpG.east)
        -- node[pos=0.86, left] {$+$} (sess.south);

    \draw[arr] (sess) -- (smain);

    \draw[arr] (2.70,1.70) -- node[pos=0.08, right] {$\Delta P_e$}
        node[pos=0.86, right] {$-$} (smain);

    \draw[arr] (smain) -- (swing);

    \draw[thick] (swing.east) -- node[above] {$\Delta \omega$} (out.center) -- (5.78,-0.62);
    \draw[arr] (5.78,0.34) -- (ieee1.east);
    \draw[arr] (5.78,-0.62) -- (ieeeG.east);

    \draw[arr] (ieee1.west) -- node[above, font=\scriptsize] {$\Delta P^1_m$} (a1.east);
    \draw[arr] (a1.west) -- node[pos=0.82, above] {$+$} (sgen.east);

    \draw[arr] (ieeeG.west) -- node[above, font=\scriptsize] {$\Delta P^G_m$} (aG.east);

    \draw[arr] (aG.west) -- (sgen |- aG.west)
        -- node[pos=0.86, left] {$+$} (sgen.south);

    \draw[arr] (sgen) -- node[pos=0.88, right] {$+$} (smain);

\end{tikzpicture}%
}
\end{center}
    \caption{Closed-loop system model with PFR and ESS support.}
    \label{fig-closed-loop}
\end{figure}
When multiple generators participate in PFR, each carries a share of the power imbalance. At steady state, standard analysis shows that the contribution of generator $i$ is
\begin{equation}\label{eq-PFRss}
    \Delta P_{\rm m,ss}^i = \frac{K^i \Delta P_{\rm e}}{\sum_{i\in \mathcal{G}_{\rm pfr}}K^i \alpha^i}, \qquad  i\in \mathcal{G}_{\rm pfr}.
\end{equation}

\subsection{Linearized Nadir Bound via Ramp Approximation}\label{NadirDerivation}
Given a frequency nadir limit $\Delta\omega_{\rm lim} < 0$, our objective is to determine the maximum electrical disturbance $\Delta P_{\rm e}$ such that $\Delta\omega(t) \geq \Delta\omega_{\rm lim}$ for all times $t$. This problem is complicated by the nonlinear SAT1 block, high-order turbine dynamics, and heterogeneous generator response rates. We proceed by describing some simplifying approximations, and then deriving the nadir bound.

First, in Figure \ref{IEEEG1-Figure} we neglect SAT2 by assuming that adequate dynamic reserves are allocated between actions. Put differently, generators do not reach their maximum output capacities when providing PFR. Second, we simplify the SAT1 block behaviour by taking into account generator redispatch during restoration. Specifically, during a power shortage the initial signal entering SAT1 is $\Delta P_{\rm ref}/T_3$. Since the timing and magnitude of load pick-ups during restoration is known, the generator references are adjusted to $\Delta P_{\rm ref}^i = \Delta P_{\rm m,ss}^i$ as defined in \eqref{eq-PFRss}. For large enough loads, this causes SAT1 to become active, and since the declining frequency from the power shortage further increases the signal entering SAT1, it will remain active when the nadir occurs. Under this approximation, the SAT1 feedback loop can be replaced by a constant source $U_{\rm o}$ entering the integrator; each generator's response is characterized by the turbine's response to a linear power ramp with slope $U_{\rm o}^i$. This is termed the \emph{ramp approximation} and is valid provided generator setpoints are chosen such that $\Delta P_{\rm ref}^i/T_3^i \geq U_{\rm o}^i$ for all $i \in \mathcal{G}_{\rm pfr}$. The block diagram in Fig.~\ref{fig-ramp-approx} depicts this approximated model.
\begin{figure}[h!]
    \centering

\begin{center}
\resizebox{0.85\columnwidth}{!}{%
\begin{tikzpicture}[auto, every node/.style={scale=0.62}]

    \tikzstyle{sum} = [draw, fill=none, circle, node distance=1cm]
    \tikzstyle{input} = [coordinate]
    \tikzstyle{output} = [coordinate]
    \tikzstyle{anch} = [coordinate]
    \tikzstyle{block} = [draw, fill=white, rectangle,
        minimum height=3em, minimum width=4em]
    \tikzstyle{wideblock} = [draw, fill=white, rectangle,
        minimum height=3em, minimum width=5.8em]
    \tikzstyle{smallblock} = [draw, fill=white, rectangle,
        minimum height=3em, minimum width=1.45em]
    \tikzstyle{satblock} = [draw, fill=white, rectangle,
        minimum height=3.2em, minimum width=1.15em,
        inner xsep=0pt, inner ysep=0pt]
    \tikzstyle{gain} = [draw, fill=white, rectangle,
        minimum height=2.1em, minimum width=2.4em]
    \tikzstyle{arr} = [thick, -{Latex[length=1.45mm,width=1.15mm]}]

    \tikzstyle{swingblock} = [draw, fill=white, rectangle,
        minimum height=3em, minimum width=4.65em]
    \tikzstyle{redblock} = [draw=red, fill=white, rectangle,
        minimum height=2.15em, minimum width=3.45em,
        inner xsep=1pt, inner ysep=0.5pt]
    \tikzstyle{intblock} = [draw=red, fill=white, rectangle,
        minimum height=2.15em, minimum width=2.4em,
        inner xsep=1pt, inner ysep=0.5pt]
    \tikzstyle{gtblock} = [draw=red, fill=white, rectangle,
        minimum height=2.15em, minimum width=3.0em,
        inner xsep=1pt, inner ysep=0.5pt]

    \node[input] (in1) at (0,1.05) {};
    \node[input] (inG) at (0,0.00) {};

    \node[block] (lp1) at (1.05,1.05)
        {$\displaystyle \frac{1}{1+s\tau}$};
    \node[block] (lpG) at (1.05,0.00)
        {$\displaystyle \frac{1}{1+s\tau}$};

    \node[sum, minimum size=0.88em, inner sep=0pt] (sess) at (2.02,1.05) {};
    \node[sum, minimum size=0.88em, inner sep=0pt] (smain) at (2.70,1.05) {};

    \node[swingblock] (swing) at (4.04,1.05)
        {$\displaystyle \frac{1}{2H_{\rm sys}s}$};

    \node[output] (out) at (6.75,1.05) {};

    \node[gain] (a1) at (3.50,0.34) {$\alpha_1$};
    \node[gain] (aG) at (3.50,-0.62) {$\alpha_G$};

    \node[sum, minimum size=0.88em, inner sep=0pt] (sgen) at (2.70,0.34) {};

    \node[gtblock] (gt1) at (4.65,0.34) {$G_{\rm T}^{1}(s)$};
    \node[intblock] (int1) at (5.65,0.34) {$\displaystyle \frac{1}{s}$};

    \node[gtblock] (gtG) at (4.65,-0.62) {$G_{\rm T}^{G}(s)$};
    \node[intblock] (intG) at (5.65,-0.62) {$\displaystyle \frac{1}{s}$};

    \node at (1.05,0.60) {$\vdots$};
    \node at (2.70,-0.14) {$\vdots$};
    \node at (3.50,-0.14) {$\vdots$};
    \node at (4.65,-0.14) {$\vdots$};
    \node at (5.65,-0.14) {$\vdots$};

    \draw[arr] (in1) -- node[pos=0.22, above] {$\Delta P^{\rm ref}_{\rm ess,1}$} (lp1);
    \draw[arr] (inG) -- node[pos=0.22, above] {$\Delta P^{\rm ref}_{\rm ess,E}$} (lpG);

    \draw[arr] (lp1) -- node[pos=0.76, above] {$+$} (sess);

    \draw[arr] (lpG.east) -- (sess |- lpG.east)
        -- node[pos=0.86, left] {$+$} (sess.south);

    \draw[arr] (sess) -- (smain);

    \draw[arr] (2.70,1.70) -- node[pos=0.08, right] {$\Delta P_e$}
        node[pos=0.86, right] {$-$} (smain);

    \draw[arr] (smain) -- (swing);

    \draw[arr] (swing.east) -- node[above] {$\Delta \omega$} (out.center);

    \draw[arr] (6.75,0.34) -- node[pos=0.12, above] {$U_0^{1}$} (int1.east);
    \draw[arr] (6.75,-0.62) -- node[pos=0.12, above] {$U_0^{G}$} (intG.east);

    \draw[arr] (int1.west) -- (gt1.east);
    \draw[arr] (gt1.west) -- node[above, font=\scriptsize] {$\Delta P^1_m$} (a1.east);
    \draw[arr] (a1.west) -- node[pos=0.82, above] {$+$} (sgen.east);

    \draw[arr] (intG.west) -- (gtG.east);
    \draw[arr] (gtG.west) -- node[above, font=\scriptsize] {$\Delta P^G_m$} (aG.east);

    \draw[arr] (aG.west) -- (sgen |- aG.west)
        -- node[pos=0.86, left] {$+$} (sgen.south);

    \draw[arr] (sgen) -- node[pos=0.88, right] {$+$} (smain);

\end{tikzpicture}%
}
\end{center}
    \caption{Open-loop system model under the ramp approximation.}
    \label{fig-ramp-approx}
\end{figure}

\textit{Nadir derivation.} Let $\mathcal{E}$ denote the set of active ESS devices. From Fig.~\ref{fig-ramp-approx}, the Laplace-domain expressions for the mechanical power output of generator $i$, the ESS power output, and the system frequency are
\begin{subequations}
\begin{align}
    \Delta P_{\rm m}^i(s) &= G_{\rm T}^i(s)\frac{U_{\rm o}^i}{s^2}, \hspace{0.4cm}
        \Delta P_{\rm s}^i(s) = \frac{\Delta P_{\rm s}^{{\rm ref},i}}{s(1+s\tau^i)}, 
        \label{eq-ess-response}\\
    \Delta\omega(s) &= \frac{1}{2H_{\rm sys}s}\big(
        \sum_{i\in\mathcal{G}_{\rm pfr}}\alpha^i\Delta P_{\rm m}^i(s)
        - \frac{\Delta P_{\rm e}}{s}
        + \sum_{i\in\mathcal{E}}\Delta P_{\rm s}^i(s)
    \big) \label{eq-xmach-w-ess}
\end{align}
\end{subequations}
The first term in \eqref{eq-xmach-w-ess} captures the combined PFR of the synchronous 
generators, the second the electrical disturbance, and the third the ESS contribution. Here

\begin{equation}\label{eq-xmach-gt}
    G_{\rm T}^i(s) = L_4^i(s)\Big(K_1^i 
        + L_5^i(s)\big(K_3^i 
        + L_6^i(s)(K_5^i + K_7^i L_7^i(s))\big)\Big)
\end{equation}
is the turbine transfer function, where $L_j(s) = (1+sT_j)^{-1}$. Replacing each $L_j(s)$ with the second-order 
polynomial approximation $L_j(s) \approx 1 - sT_j + s^2T_j^2$ (accurate for 
frequencies up to $1/T_j$~rad/s) and neglecting terms of order higher than two 
in $s$ yields
\begin{equation}\label{eq-xmach-ca}
    G_{\rm T}^i(s) \approx c_1^i - c_2^i s + c_3^i s^2,
\end{equation}
where the coefficients $c_1^i, c_2^i, c_3^i$ depend only on turbine parameters 
and are tabulated in~\cite{HarryThesis}. Similarly, retaining first-order terms 
in the ESS response gives
\begin{equation}\label{eq-ess-response-approx}
    \Delta P_{\rm s}^i(s) \approx \frac{(1-s\tau^i)\Delta P_{\rm s}^{{\rm ref},i}}{s}.
\end{equation}

Substituting \eqref{eq-xmach-ca}--\eqref{eq-ess-response-approx} into \eqref{eq-xmach-w-ess} and collecting terms by power of $s$ yields
\begin{equation}\label{eq-xmach-ramp-laplace-ess}
    \Delta\omega(s) \approx \frac{1}{2H_{\rm sys}s}\!\left(
        \frac{C_1}{s^2}
        - \frac{C_2 + \Delta P_{\rm e} - P_{\rm s}^{\rm tot}}{s}
        + C_3 - P_{\rm s}^{\tau}
    \right),
\end{equation}
where the aggregate generator response coefficients are
\begin{equation} \label{eq-xmach-C}
    C_1 = \sum_{i\in\mathcal{G}_{\rm pfr}}\alpha^i U_{\rm o}^i, C_2 = \sum_{i\in\mathcal{G}_{\rm pfr}}\alpha^i U_{\rm o}^i c_2^i, C_3 = \sum_{i\in\mathcal{G}_{\rm pfr}}\alpha^i U_{\rm o}^i c_3^i, 
\end{equation}
and the aggregate ESS contributions are
\begin{equation}\label{eq-ess-agg}
    P_{\rm s}^{\rm tot} = \sum_{i\in\mathcal{E}}\nolimits\Delta P_{\rm s}^{{\rm ref},i}, \qquad P_{\rm s}^{\tau}    = \sum_{i\in\mathcal{E}}\nolimits\tau^i\Delta P_{\rm s}^{{\rm ref},i}
\end{equation}
with $P_{\rm s}^{\rm tot}$ the total ESS setpoint change and $P_{\rm s}^{\tau}$ the 
response-time-weighted ESS contribution. Inverting \eqref{eq-xmach-ramp-laplace-ess} 
gives the time-domain frequency deviation
%
%
\begin{equation}\label{eq-xmach-time-domain}
    \Delta\omega(t) = \frac{1}{2H_{\rm sys}}\!\left(
        \frac{C_1 t^2}{2}
        - \bigl(C_2 + \Delta P_{\rm e} - P_{\rm s}^{\rm tot}\bigr)t
        + C_3 - P_{\rm s}^{\tau}
    \right)
\end{equation}
%
%
This is a convex quadratic in $t$, with minimum at 
\begin{equation}
    t_{\rm nad} = \frac{C_2 + \Delta P_{\rm e} - P_{\rm s}^{\rm tot}}{C_1},
\end{equation}
giving the frequency nadir approximation
\begin{equation} \label{eq-xmach-wnadir-ess}
    \Delta\omega_{\rm nad}
    \define \frac{1}{2H_{\rm sys}}\!\left(
        C_3 - P_{\rm s}^{\tau}
        -\frac{\bigl(C_2 + \Delta P_{\rm e} - P_{\rm s}^{\rm tot}\bigr)^2}{2C_1}
    \right).
\end{equation}
Enforcing $\Delta\omega_{\rm nad} \geq \Delta\omega_{\rm lim}$ and solving for $\Delta P_{\rm e}$ yields the nonlinear bound \eqref{eq-ess-PLmax}. Since the ESS setpoints $\Delta P_{\rm s}^{{\rm ref},i}$ are decision variables in the MILP, a first-order Taylor expansion around $\Delta P_{\rm s}^{{\rm ref},i} = 0$ linearizes the bound, and we obtain the linear bound
\begin{equation}\label{eq-ess-PLmax-linear}
    \Delta P_{\rm e} \leq \Delta P_{\rm e,max}
    \triangleq g_{0} + \sum_{i\in\mathcal{E}}\ g_{\rm s}^{i}\Delta P_{\rm s}^{{\rm ref},i},
\end{equation}
where
\begin{subequations} \label{eq-g-coeffs}
\begin{align}
    g_{0} &\triangleq \sqrt{4H_{\rm sys}C_1|\Delta\omega_{\rm lim}| + 2C_1C_3} - C_2, \label{eq-g0}\\
    g_{\rm s}^i &\triangleq 1 - \frac{C_1\tau^i}{\sqrt{4H_{\rm sys}C_1|\Delta\omega_{\rm lim}| + 2C_1C_3}}.\label{eq-gis}
\end{align}
\end{subequations}
Here $g_{0}$ is interpreted as the maximum tolerable load pickup without ESS support, and $g_{\rm s}^i \in [0,1]$ quantifies the marginal benefit of ESS $i$: faster ESSs (smaller $\tau^i$) yield larger $g_{\rm s}^i$ and thus enable greater load pickup. The framework also extends to systems with nonzero load damping, see \cite{HarryThesis, EPEC2025}.
\begin{figure*}[!t]
\begin{equation} \label{eq-ess-PLmax}
    \Delta P_{\rm e} \leq
    \sqrt{4H_{\rm sys}C_1|\Delta\omega_{\rm lim}|
        + 2C_1C_3
        - 2C_1\sum_{i\in\mathcal{E}}\tau^i\Delta P_{\rm s}^{{\rm ref},i}}
    - C_2 + \sum_{i\in\mathcal{E}}\Delta P_{\rm s}^{{\rm ref},i}
\end{equation}
\end{figure*}

\section{Frequency-Constrained MILP for Dynamic Restoration Planning}\label{sec-frequency-cons}

Our goal is now to integrate the frequency nadir constraint \eqref{eq-ess-PLmax-linear} into the MILP of Section \ref{sec-Freq_const_MILP}. During restoration, net power imbalances result from load pick-ups, generator cranking, and ESS charging or discharging. Thus, the left-hand side of \eqref{eq-ess-PLmax-linear} at each step can be written as a linear expression of status variables as
\begin{equation}\label{eq-pedef}
\begin{aligned}
    \boldsymbol{\Delta P}_{\rm e}[k]
    &= \boldsymbol{P}_{\rm d}^{\top}\bigl(\boldsymbol{b}_{\rm d}[k] 
        - \boldsymbol{b}_{\rm d}[k-1]\bigr)\\
     &\quad +\boldsymbol{P}_{\rm c}^{\top}\bigl(\boldsymbol{b}_{\rm gc}[k] 
        - \boldsymbol{b}_{\rm gc}[k-1]\bigr),
\end{aligned}
\end{equation}
where $\boldsymbol{P}_{\rm d}, \boldsymbol{P}_{\rm c}$ are the fixed vectors of load demand and cranking powers. The right-hand side of \eqref{eq-ess-PLmax-linear} requires knowledge of the system inertia and the PFR coefficients at each step $k$. However, these parameters depend on which
generators are online or ramping at that step. Using the generator
phase variables $\boldsymbol{b}_{\rm gr}[k]$ and
$\boldsymbol{b}_{\rm go}[k]$ introduced in
Section~\ref{NBSU-Start}, the system inertia constant and the
participation factor vector at step $k$ can be expressed as
\begin{align}
    \boldsymbol{H}_{\rm sys}[k]
        = \overline{P}&_{\rm g}^{\top}
        \Bigl(H \odot \bigl(\boldsymbol{b}_{\rm gr}[k]
        + \boldsymbol{b}_{\rm go}[k]\bigr)\Bigr),
        \label{eq-Hsys-vec}\\
    \boldsymbol{\alpha}[k]
        &= \overline{P}_{\rm g} \odot \boldsymbol{b}_{\rm go}[k],
        \label{eq-alpha-vec2}
\end{align}
where $H \in \mathbb{R}^{G}$ contains the per-generator inertia constants. The aggregate PFR
coefficients $\boldsymbol{C_1}[k]$, $\boldsymbol{C_2}[k]$, $\boldsymbol{C_3}[k]$ follow
from $\boldsymbol{\alpha}[k]$ via~\eqref{eq-xmach-C}, and the coefficients of the power imbalance bound $\boldsymbol{g}_{\rm 0}[k] \in \mathbb{R}$ and $\boldsymbol{g}_{\rm s}[k] \in \mathbb{R}^{S}$ follow by substitution in \eqref{eq-g-coeffs}. The nadir-enforcing
bound~\eqref{eq-ess-PLmax-linear} at step $k$ becomes
\begin{equation} \label{eq-ess-milp}
    \Delta \boldsymbol{P}_{\rm e}[k] \leq \boldsymbol{g}_0[k]
    + \boldsymbol{g}_{\rm s}[k]^{\top}
    \Delta \boldsymbol{P}_{\rm s}^{\rm ref}[k],
\end{equation}
where $\Delta \boldsymbol{P}_{\rm e}[k]$ is given by \eqref{eq-pedef} and  $\boldsymbol{\Delta P}_{\rm s}^{\rm ref}[k] \in \mathbb{R}^{S}$ is
the vector of ESS power setpoint changes at step $k$. 
Due to \eqref{eq-g-coeffs}, the coefficients $\boldsymbol{g}_0[k]$ and $\boldsymbol{g}_{\rm s}[k]$ are nonlinear functions of the binary generator status variables, and thus \eqref{eq-ess-milp} is a nonlinear constraint. 

The nadir-constrained restoration problem including \eqref{eq-ess-milp} is therefore a mixed-integer
nonlinear program~(MINLP), which is computationally intractable at scale. The rolling-horizon algorithm presented in Section~\ref{Iterative-Rolling} resolves this by freezing these coefficients at each iteration, converting the MINLP into a sequence of tractable MILPs.

\subsection{Rolling-Horizon Algorithm: From MINLP to MILP}
\label{Iterative-Rolling}

The full nadir-constrained restoration planning problem can be written in a compact abstract form as
\begin{equation}\label{eq-minlp}
    \min_{x}\; f(x)
    \quad \text{s.t.} \quad
    h(x,\,\mu(x)) \leq 0,
\end{equation}
where $x$ collects all binary and continuous decision variables over all time
(network element statuses, generator and ESS setpoints,
power flow variables), and $f(x)$ is the linear restoration objective
function from~\eqref{eq-milp-obj}. 
The term $h(x,\mu(x)) \leq 0$ represents all constraints, including the frequency nadir constraint~\eqref{eq-ess-milp}; the parameter $\mu(x) = (\boldsymbol{g}_{\rm 0}(x), \boldsymbol{g}_{\rm s}(x))$ denotes the parameters in the expression \eqref{eq-ess-milp}, which are themselves (nonlinear) functions of $x$. While for any fixed $\mu$, $x \mapsto h(x,\mu)$ is a linear function, the problem \eqref{eq-minlp} is an intractable MINLP.

\begin{figure}[h!]
\centering
\includegraphics[width=0.8\linewidth]{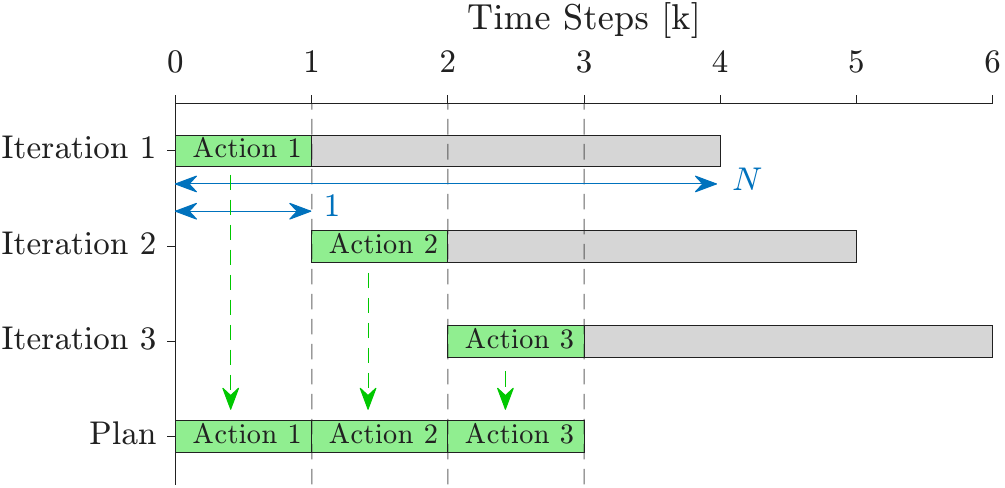}
\caption{Rolling-horizon optimization restoration planning. At each iteration $\nu$,
         the MILP is solved over a look-ahead window. The actions in the committed period (green) are fixed to the plan; the remaining actions (gray) are discarded. The window then advances by $N$ steps and the parameters $\mu$ are recomputed.}
\label{fig-receding}
\end{figure}

We resolve this nonlinearity by solving the problem iteratively in an $N$-step receding horizon manner,
where at each iteration the parameters $\mu(x)$ are
updated using the current and previous restoration states, and then fixed
as constants in \eqref{eq-minlp}. This reduces the problem into solving a series of MILPs. The full procedure is summarized in
Algorithm~\ref{alg-irsc-pu} and illustrated in
Fig.~\ref{fig-receding}, and will now be explained.

\begin{algorithm}
\caption{Rolling-Horizon Restoration Planning}
\label{alg-irsc-pu}
\begin{algorithmic}[1]
\State $x_0 \leftarrow$ initial restoration state;\quad
       $\mathsf{ResPlan} \leftarrow [x_0]$;\quad $\nu \leftarrow 1$
\While{restoration incomplete}
    \State Compute and fix $\hat{\mu}^{(\nu)}$
           via \eqref{eq-g-coeffs} using $\mathsf{ResPlan}$
    \State Solve MILP \eqref{eq-minlp} with fixed $\hat{\mu}^{(\nu)}$ to obtain $x^{(\nu)}$
    \State Save first state:\;
           append $x^{(\nu)}[1]$ to $\mathsf{ResPlan}$
    \State $\nu \leftarrow \nu+1$
\EndWhile
\State \textbf{return} $\mathsf{ResPlan}$
\end{algorithmic}
\end{algorithm}

In line 3 of Algorithm~\ref{alg-irsc-pu} at iteration $\nu$, an $N$-step \emph{prediction} $\hat{\mu}^{(\nu)}$ of $\mu(x)$ is computed from $\mathsf{ResPlan}$ as follows. Since all generator cranking and ramping times are known, the phase variables
$\boldsymbol{b}_{\rm gr}[k]$ and $\boldsymbol{b}_{\rm go}[k]$ depend solely on when each generator is started. If we assume that no additional generators are activated beyond those already committed, these variables can be predicted for all future times $k \in \{1,\ldots,N\}$ via the expressions~\eqref{eq-milp-aux}, using information encoded in $\mathsf{ResPlan}$. 
We can then form predictions of inertias 
$\hat{{\boldsymbol{H}}}_{\rm sys}^{(\nu)}[k]$, participation factors
$\hat{{\boldsymbol{\alpha}}}^{(\nu)}[k]$, and PFR coefficients
$\hat{\boldsymbol{C}}_1^{(\nu)}[k]$, $\hat{\boldsymbol{C}}_2^{(\nu)}[k]$,
$\hat{\boldsymbol{C}}_3^{(\nu)}[k]$, and subsequently predictions of $ \hat{{\boldsymbol{g}}}_{\rm 0}^{(\nu)}[k]$ and $\hat{{\boldsymbol{g}}}_{\rm s}^{(\nu)}[k]$ from \eqref{eq-g-coeffs}.

In Step 4 of Algorithm \ref{alg-irsc-pu}, the predicted parameter values
\begin{equation}
\hat{\mu}^{(\nu)} = \bigl\{ 
        \hat{{\boldsymbol{g}}}_{\rm 0}^{(\nu)}[k],\hat{{\boldsymbol{g}}}_{\rm s}^{(\nu)}[k]
        \bigr\}_{k=1}^{N}
\end{equation}
are inputs in the optimization problem \eqref{eq-minlp}, and no longer depend on the optimization variable $x$. In particular, the nonlinear
constraint~\eqref{eq-ess-milp} becomes the linear constraint 
\begin{equation}\label{eq-ess-milp-frozen}
    \Delta \boldsymbol{P}_{\rm e}[k] \leq
    \hat{{\boldsymbol{g}}}_{\rm 0}^{(\nu)}[k]
    + \hat{{\boldsymbol{g}}}_{\rm s}^{(\nu)}[k]^{\top}
    {\Delta \boldsymbol{P}}_{\rm s}^{\rm ref}[k].
\end{equation}
The MINLP~\eqref{eq-minlp} reduces to a standard MILP at
iteration $\nu$, which is solved for an $N$-step plan $x^{(\nu)}=(x^{(\nu)}[1],\dots,x^{(\nu)}[N])$. Only the first step of the solution, $x^{(\nu)}[1]$, is committed to the plan in each iteration, and the rest are discarded. This approach allows the predicted parameters $\hat{\mu}^{(\nu)}$ to be updated as restoration planning progresses, and the process repeats until all network elements are restored.

Regarding the constraint \eqref{eq-ess-milp-frozen}, 
as restoration progresses and more generators synchronize, the 
inertia $\hat{{\boldsymbol{H}}}_{\rm sys}^{(\nu)}[k]$ grows, which increases the
upper bound in \eqref{eq-ess-milp-frozen}
and permits larger load pick-ups while maintaining frequency security. This provides
a principled, physics-based alternative to more static operator heuristics
such as the IESO 5\% rule~\cite{IESO2024}.

\section{Case Study}\label{CaseStudy}
\begin{figure}[h!]
    \centering
    \includegraphics[width=0.7\linewidth]{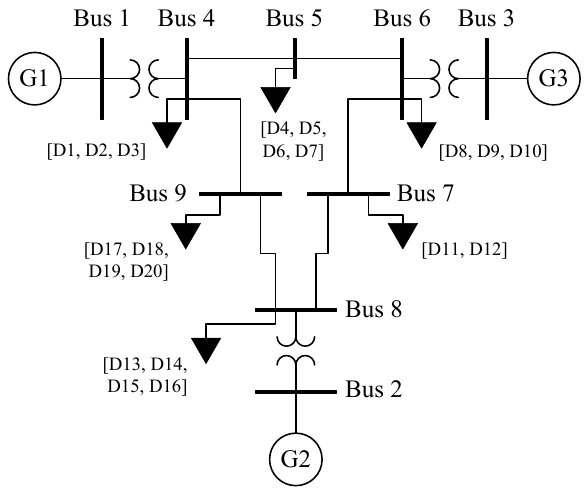}
    \caption{Modified IEEE 9-bus system. Numbers in brackets 
    denote the indices of loads on each bus.}
    \label{Case9-figure}
\end{figure}

The proposed frequency-constrained restoration framework with ESS coordination is tested on the modified IEEE 9-bus system shown in Fig.~\ref{Case9-figure}. Generator parameters and IEEEG1 turbine-governor models follow~\cite{Anderson2003,Birchfield2022,Neplan2015}. Loads are divided into blocks, ranging between 3--16~MW. Generator~1 is designated as the BSU and generators~2--3 as NBSUs. Additional system parameters are provided in~\cite{EPEC2025,HarryThesis}.

We compare three restoration plans: (a)~\textit{no frequency constraint}, where the MILP \eqref{Eq:TrueMILP} is solved without the frequency constraint~\eqref{eq-ess-milp}; (b)~\textit{5\% rule-of-thumb}, where \eqref{Eq:TrueMILP} is solved subject to the additional linear constraint
\begin{equation}\label{eq-ieso-rule}
    \Delta P_{\rm e}[k] \leq 0.05\, \overline{P}_{\rm g}^{\top} 
    \boldsymbol{b}_{\rm go}[k],
\end{equation}
which states that the power imbalance at each step is limited to 5\% of the total online generation capacity~\cite{IESO2024}, and finally (c)~\textit{proposed framework} as described in in Algorithm~\ref{alg-irsc-pu}, with frequency bound~\eqref{eq-ess-milp} with 
$\Delta\omega_{\rm lim} = -1$~Hz. In all cases, MILPs are formulated using YALMIP~\cite{Lofberg2004} and solved with Gurobi~\cite{gurobi}. Restorative actions are performed every two minutes to allow transient dynamics to settle between steps.

The different plans are compared through two dynamic simulations: (i)~a MATLAB ODE simulation of the closed loop ASF model, which has $6G+1$ states with the center-of-inertia frequency deviation as the primary state of interest, with inputs $\boldsymbol{\Delta P}_{\rm e}$ and $\Delta\boldsymbol{P}_{\rm ref}$ at each step; and (ii) for the proposed method, a full PSS/E dynamic simulation, which captures transient dynamics, inter-machine interactions, and other details neglected by the ASF model.


\subsection{Comparison of Restoration Plans without ESS}\label{Freq-Res-Seq}

\begin{figure}[h!]
    \centering
    \includegraphics[width=.9\linewidth]{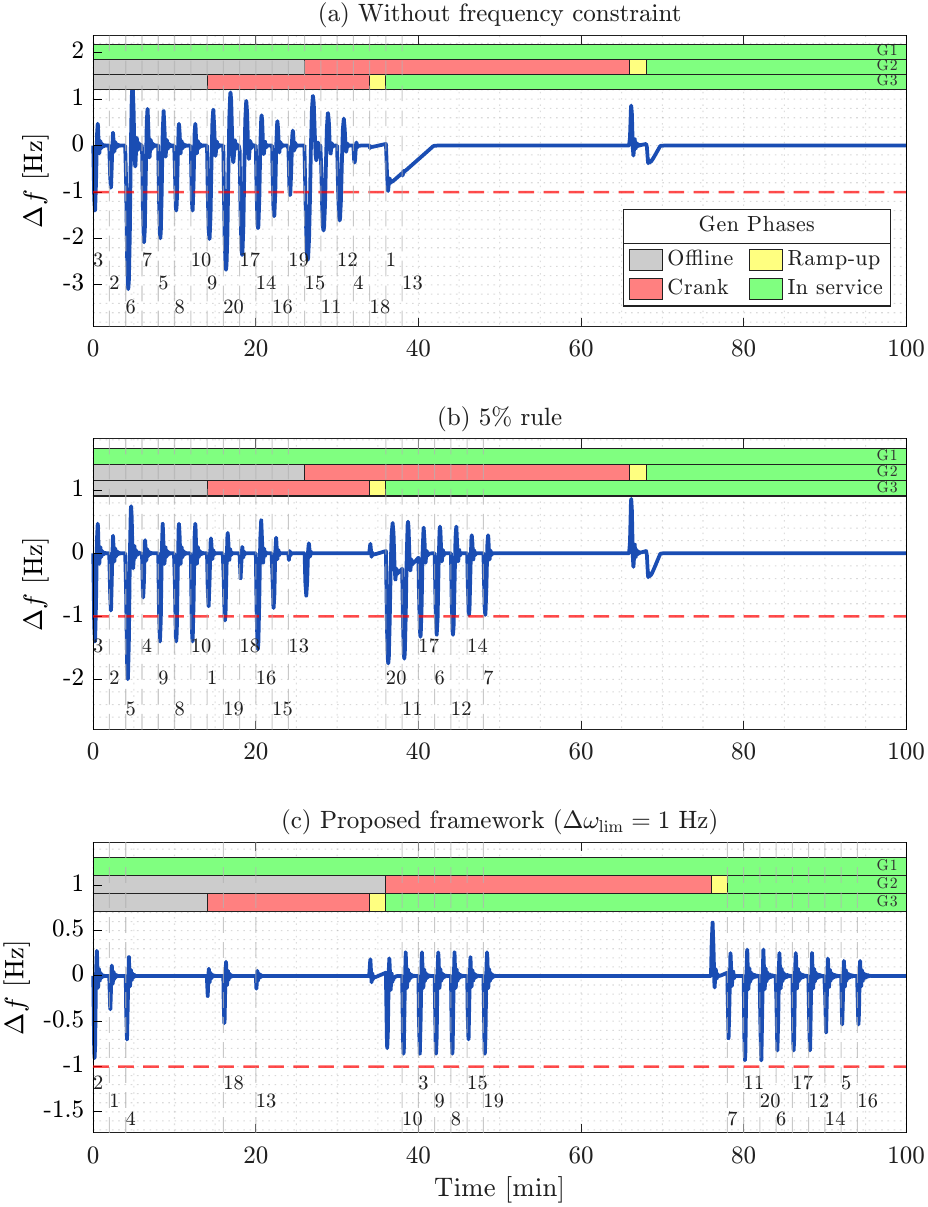}
    \caption{Center-of-inertia frequency profiles (MATLAB ODE simulation) for the three restoration plans. Numbers indicate indices of restored loads.}
    \label{Freq-stacked-noESS}
\end{figure}

We first examine the case without ESS support. Fig.~\ref{Freq-stacked-noESS} shows the center-of-inertia frequency profiles from MATLAB simulation for all three restoration plans. 
Without frequency limits, the algorithm energizes generators and loads in rapid succession, as shown in Fig.~\ref{Freq-stacked-noESS}-(a). The resulting plan experiences severe frequency excursions exceeding $-3$~Hz due to the early energization of large loads before adequate generation capacity and inertia are available. In practice, such deviations would trigger underfrequency protection relays, risking further system failures that may halt or compromise the restoration effort. This plan is the fastest of the three sequences, but is operationally infeasible.


In Fig.~\ref{Freq-stacked-noESS}-(b), plan (b) prevents the most severe frequency excursions from the first plan. However, the load pick-up limit \eqref{eq-ieso-rule} does not accurately account for the actual system inertia and PFR capability at each step. Consequently, frequency deviations as large as $-2$~Hz still appear.
Static limits can only be effective if they are specifically tuned to the system in question. In this case, the limit fails to enforce the desired frequency bound, while in other cases, it may be overly conservative and slow the restoration process.

\begin{figure}[h!]
    \centering
    \includegraphics[width=1\linewidth]{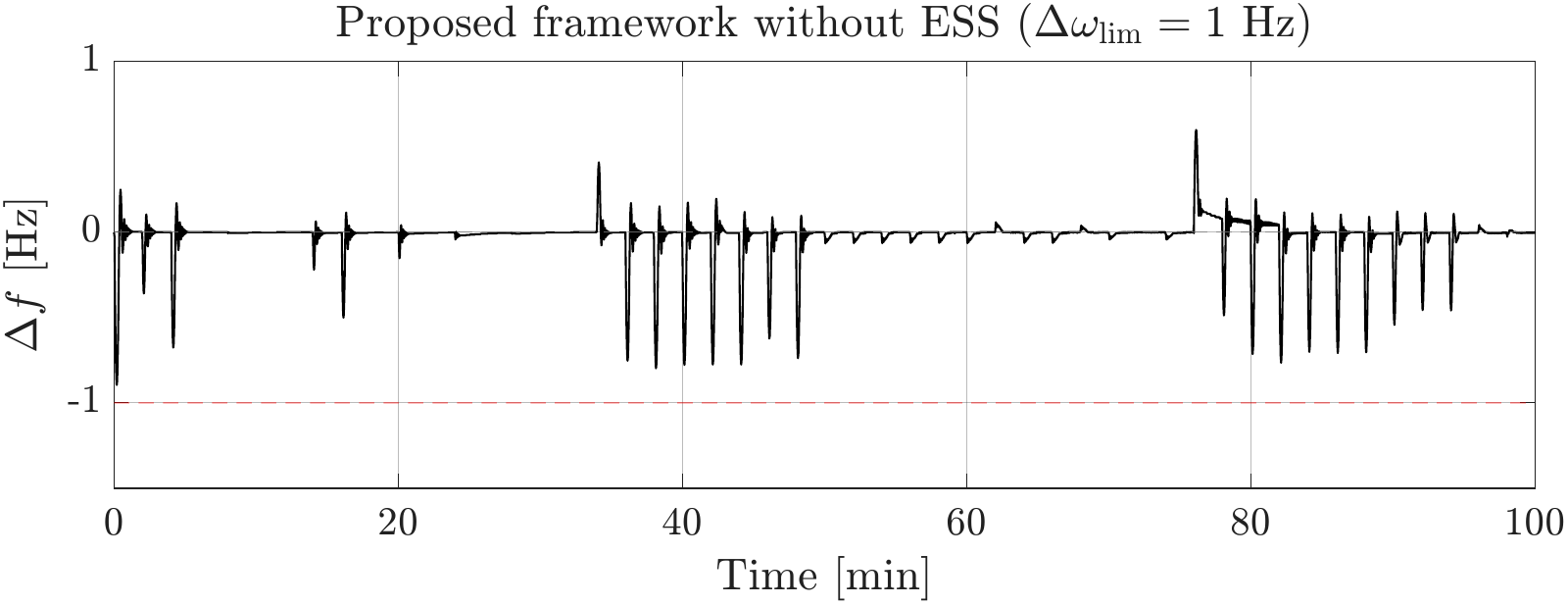}
    \caption{Center-of-inertia frequency profiles (PSS/E simulation) for the~proposed framework.}
    \label{fig-PSSE-noESS}
\end{figure}

The proposed frequency nadir constraint~\eqref{eq-ess-milp} determines the maximum allowable load pickup based on the actual system inertia and PFR capacity at each step. The plan obtained via Algorithm~\ref{alg-irsc-pu} is shown in Fig.~\ref{Freq-stacked-noESS}-(c). During the early stages, when the system has low inertia, only small loads are energized. As NBSUs become synchronized, the allowable imbalance (RHS of \eqref{eq-ess-milp}) increases and larger loads can be safely picked up while keeping the frequency within the 1~Hz limit. 
Fig.~\ref{fig-PSSE-noESS} presents the center-of-inertia frequency profiles from PSS/E simulation using the same plan from the proposed framework. The results are consistent with Fig.~\ref{Freq-stacked-noESS}-(c)\textemdash the frequency bound is satisfied during the entire sequence.
This PSS/E simulation provides evidence that our approach remains accurate in much more detailed dynamic simulation settings.

\subsection{Proposed Framework with ESS Participation}\label{ESS-Participate}
\begin{figure}[h!]
\centerline{\includegraphics[width=1\linewidth]{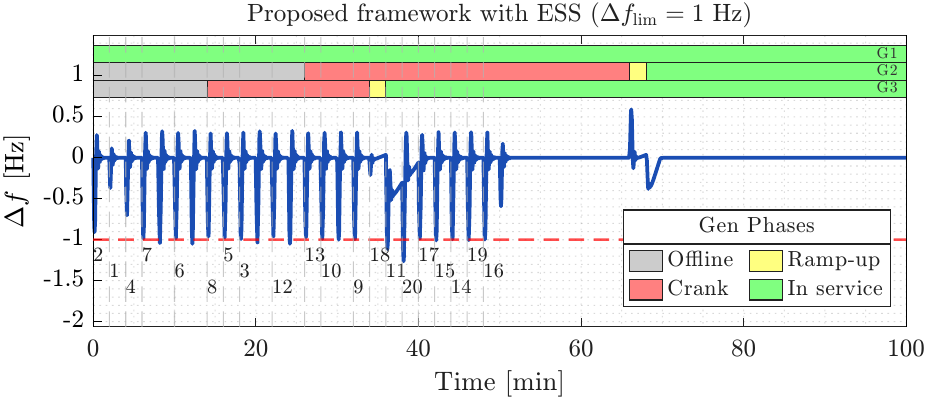}}
    \caption{Center-of-inertia frequency profiles (MATLAB ODE simulation) for the proposed restoration framework with ESS coordination. Numbers indicate the indices of loads restored at each step.}
    \label{Freq-ESS}
\end{figure}

In this scenario, an ESS unit (50/10 MWh/MW) is connected at bus~5, and a new plan using the proposed frequency constraint \eqref{eq-ess-milp-frozen} is recomputed via Algorithm \ref{alg-irsc-pu}.
Fig.~\ref{Freq-ESS} shows the  MATLAB ASF simulations of the system's frequency deviation during the execution of the restoration plan.

In the plan without ESS support (Fig.~\ref{Freq-stacked-noESS}(c)), larger loads must wait until additional NBSUs are synchronized and the system has sufficient inertia to safely energize them. In contrast, in Fig.~\ref{Freq-ESS} the ESS charging and discharging actions are coordinated with generation and load energization to help offset the resulting power imbalances. The ESS effectively compensates for low system inertia during early restoration stages, when few generators have been synchronized, reducing the total restoration time from 94 to 50 minutes. The frequency dips that slightly exceed the predefined limit are caused by the linearization step in \eqref{eq-ess-PLmax-linear}, which results in optimistic nadir predictions when ESSs are dispatched. If strict satisfaction is required, the limit $\Delta \omega_{\rm lim}$ can be tightened.

\begin{figure}[h!]
    \centering
    \includegraphics[width=1\linewidth]{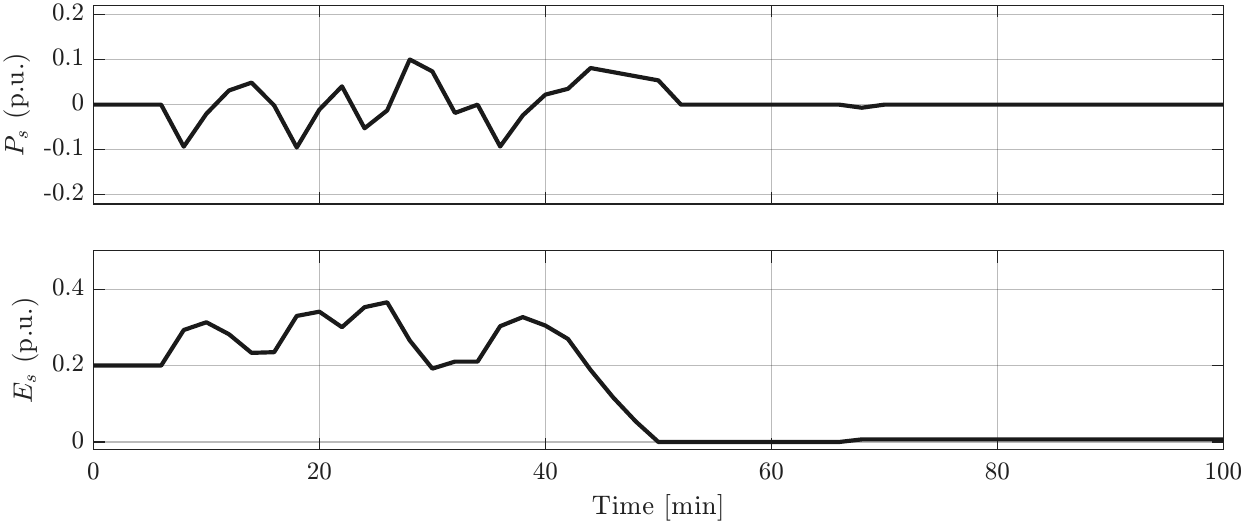}
    \caption{ESS power setpoints and state of charge during restoration under  plans (c).}
    \label{Case9-ESS}
\end{figure}

The ESS charging and discharging actions can be observed in Fig.~\ref{Case9-ESS}, which shows the ESS power setpoints and state of charge throughout the restoration sequence. Together, Fig.~\ref{Freq-ESS} and Fig.~\ref{Case9-ESS} demonstrate that strategic charging/discharging during restoration can significantly decrease restoration time. 

\begin{figure}[h!]
    \centering
    \includegraphics[width=1\linewidth]{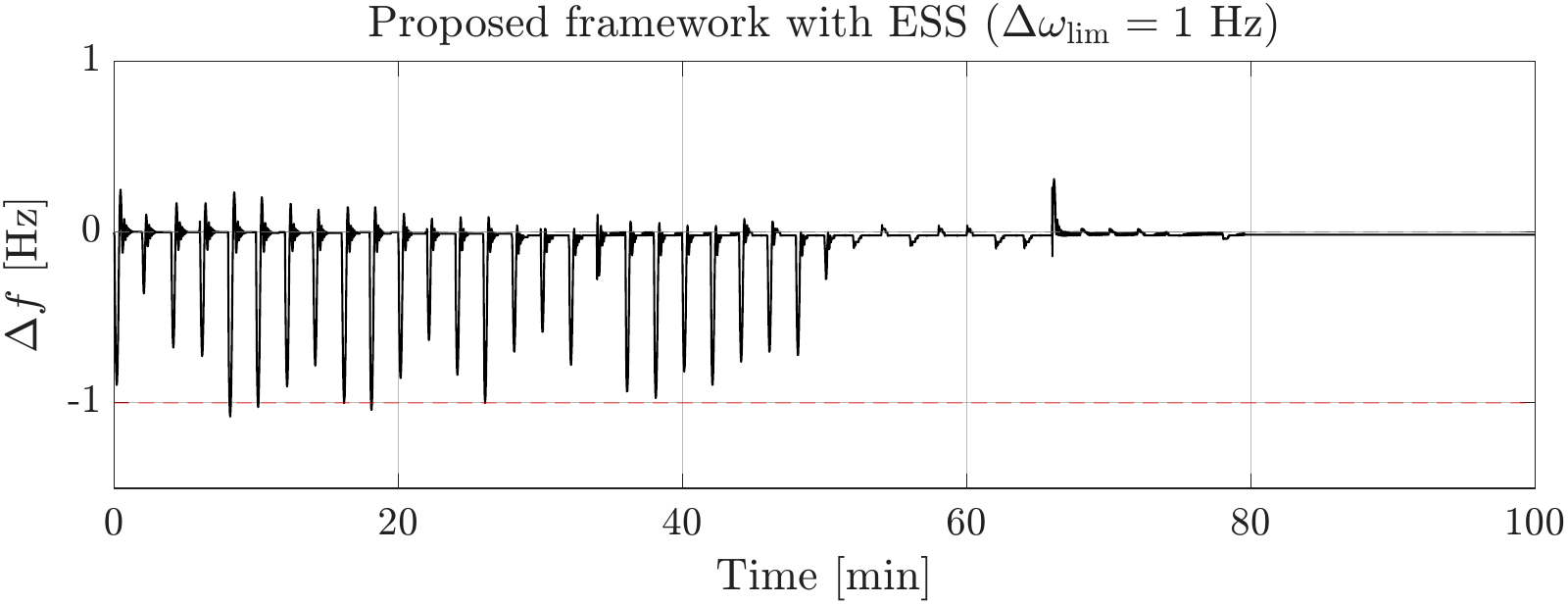}
    \caption{Center-of-inertia frequency profiles (PSS/E simulation) 
    for~proposed framework with ESS coordination.}
    \label{fig-psse-ess}
\end{figure}

To validate these results in a high-fidelity dynamic simulation environment, the same restoration plan was simulated in PSS/E; the corresponding center-of-inertia frequency profile is plotted in Fig.~\ref{fig-psse-ess}. The traces are consistent with the MATLAB ODE results in Fig.~\ref{Freq-ESS}, again highlighting that even simple dynamic models can produce benefits when properly integrated into restoration planning.

\section{Conclusion}\label{Conclusion}
We developed and validated frequency-constrained MILP framework for transmission system restoration
with ESS coordination. The frequency nadir constraints address limitations of existing methods that rely on heuristics, which may be conservative or optimistic in practice.
The integration of frequency nadir limits, derived based on IEEEG1 governor-turbine and first-order ESS dynamics, ensure the algorithm only produces safe restoration plans that do not result in under-frequency violations.
Case studies on a modified IEEE 9-bus system demonstrate that the algorithm generates frequency-secure restoration plans, as validated through dynamic simulations in PSS/E. Furthermore, ESS integration can significantly reduce restoration time in frequency-secure plans, providing both power and frequency support through setpoint adjustments.
The framework provides a systematic and computationally efficient approach to generate frequency-secure restoration plans, offering actionable insights for system operators.
Future work will extend the framework to include AC power flow relaxations, voltage constraints, and multiple islands.

\bibliographystyle{IEEEtran}
\bibliography{brevalias,References}

\end{document}